# JGR Space Physics

# Patch Size Evolution During Pulsating Aurora

Noora Partamies[1,2] 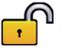, Karl Bolmgren[3,4] , Erkka Heino[1,5] , Nickolay Ivchenko[4,6] , Joseph E. Borovsky[7] , and Hanna Sundberg[4,8]

[1]Department of Arctic Geophysics, The University Centre in Svalbard, Longyearbyen, Norway, [2]Birkeland Centre for Space Science, Bergen, Norway, [3]Department of Electronic and Electrical Engineering, University of Bath, Bath, UK, [4]School of Electrical Engineering, KTH, Royal Institute of Technology, Stockholm, Sweden, [5]Department of Physics and Technology, UiT The Arctic University of Norway, Tromsø, Norway, [6]South African National Space Agency, Hermanus, South Africa, [7]Space Science Institute, Boulder, CO, USA, [8]Now at FOI Swedish Defence Research Agency, Stockholm, Sweden

**Abstract** We report both decreasing and increasing trends in the patch sizes during pulsating aurora events. About 150 pulsating auroral events over the Fennoscandian Lapland have been successfully analyzed for their average patch size, total patch area, and number of patches as a function of event time, typically 1–2 hr. An automatic routine has been developed to detect patches in the all-sky camera images. In addition to events with decreasing and increasing average patch size evolution over the course of the pulsating aurora, events with no size trends and events with intermittently increasing and decreasing patch size trends were also found. In this study, we have analyzed a subset of events for which the average and total patch size systematically increase or decrease. The events with increasing patch size trend do not experience a decrease in the peak emission height, which was previously associated with the behavior of pulsating aurora precipitation. Furthermore, the events with increasing patch sizes have shorter lifetimes and twice as many substorm-injected energetic electrons at geosynchronous orbit as the events with decreasing patch sizes. Half of the events with increasing patch sizes occur during substorm expansion phases, while a majority (64%) of the ones with decreasing patch sizes take place during the recovery phase. These findings suggest that the visual appearance of pulsating aurora may be used as an indication of the pulsating aurora energy deposition to the atmosphere.

## 1. Introduction

Pulsating aurora (PsA) is described by regions of weak emission turning on and off with a period of about 2–20 s (Lessard, 2012) or, maybe more precisely, emission regions whose intensity varies between states of dim and bright (Grono & Donovan, 2018). The emission regions can appear in different shapes from elongated arc-like structures to irregular patches in very different sizes. The periodicity can be irregular even for a single structure (Humberset et al., 2016), and different periods often take place within an observer's field of view (FOV) simultaneously. PsA occurs most frequently during substorm recovery phases and in the morning sector. The most likely mechanism for PsA precipitation is concluded to be wave-particle interaction due to chorus waves at the equatorial magnetosphere (Kasahara et al., 2018; Nishimura et al., 2018). Chorus wave activity is triggered by cyclotron resonant instability with plasma sheet electrons, which are injected to the inner magnetosphere during geomagnetic activity. Therefore, this type of precipitation typically occurs at the equatorward part of the auroral oval (Thorne et al., 2010).

Based on the European Incoherent Scatter radar observations of *E* region peak electron density and the corresponding height, Hosokawa and Ogawa (2015) found the electron density peak height to be lower during the PsA as compared to other aurora. At the same time with lower electron density peak heights, higher electron density values were recorded. In addition, their results showed an anticorrelation between the electron density peak heights and auroral electrojet (AE) index values during substorm activity prior to the PsA.

Electron density enhancements during PsA events have been observed to reach down to about 68 km, which is in agreement with precipitation energies up to about 200 keV (Miyoshi et al., 2015). Lower band chorus waves observed simultaneously by the Van Allen Probe-A satellite were modeled to produce a wide range of precipitation energies by scattering both relativistic (up to megaelectron volt) and auroral (of the order of kiloelectron volts) electrons. This model calculation was supported by Grandin et al. (2017) who reported a nearly one-to-one correlation between the temporal evolution of optical emission intensity and that of cosmic radio noise absorption (CNA). As auroral emission and CNA are driven by different precipitation energies, the high correlation between them suggests that the pulsation modulation takes place over a large range of energies rather than the high-energy tail alone. Precipitation energies over 10 keV during PsA have also been observed to modulate the close range coherent echoes of the Super Dual Auroral Radar Network radars in a good correlation with CNA (Milan et al., 2008). However, the CNA values during auroral pulsations are typically low (at the level of about 0.5 dB) indicating that the fluxes of high-energy particles also tend to be low, as compared to substorm absorption (>1 dB), for instance.

European Incoherent Scatter radar observations of electron densities during PsA have further showed an electron density peak height variation between 93 and 100 km (Oyama et al., 2016). Auroral peak emission heights determined by stereoscopic auroral imaging (Kataoka et al., 2016) resulted in even lower values of 85–95 km during PsA, while simultaneous discrete aurora were estimated to occur at and above 100 km. Reports on enhanced electron density and low peak emission have inspired studies of chemical changes in the mesospheric composition due to strong ionization. Using a precipitating electron source with energies up to about 200 keV, Turunen et al. (2016) modeled the ionospheric chemistry for PsA. After 30 min of PsA the model calculations showed a depletion in mesospheric odd oxygen of the order of several tens of percent. This depletion was accompanied by a 10–20% increases in $NO_x$ and $HO_x$ species at and below the mesopause region. These results suggest that energetic electrons during PsA may cause noticeable changes in the neutral atmosphere.

In a recent statistical study of about 400 PsA events (Partamies et al., 2017) the peak emission height of the aurora was observed to experience an average decrease by about 8 km at the start of the events. This brings the median heights down to about 107 km and the lowest 25% of the events down to about 90 km. This systematic decrease of peak emission heights is interpreted as a proxy for hardening of the precipitation throughout the precipitation spectrum, although the direct evidence only includes energies, which produce optical emissions. As illustrated by an example event in Figure 3 of Partamies et al. (2017), the patches often become smaller toward the end of the event. However, also events with growing patches have been identified, which raises a question of the involvement of the precipitation energy as a controlling factor for the patch size. Two commonly used precipitation energy proxies, an emission ratio and auroral peak emission height, did not show a clear distinction between the events with decreasing or increasing patch sizes (Bolmgren, 2017). The aim of the current study is to examine the dependence of patch size on the energy deposition during the PsA events by using more global measures. A visual indication of particle precipitation energies during the PsA may become helpful in judging the importance of these events to the mesospheric chemistry.

## 2. Ground-Based Auroral Observations

In 1997–2007 the MIRACLE network of all-sky cameras (ASC) included five identical instruments (Sangalli et al., 2011; Syrjäsuo et al., 1998) in the Fennoscandian Lapland with overlapping FoVs, as shown by map in Figure 1. The common imaging mode includes a green filtered image at 557.7 nm every 20 s and blue (427.8 nm) and red (630.0 nm) filtered images once a minute. A 1-s exposure time is typically used for green emission, while that for red and blue emission is 2 s. The imaging season in Lapland lasts from September until March. Images of the night sky are automatically captured whenever the Sun is more than 10° below the horizon. This results in about 0.8 million images per year per station. These data are pruned to leave only images, which include aurora, for further analysis. More details of the camera system, imaging, and the pruning procedure can be found in, for example, Partamies et al. (2014).

The event selection has been described in more detail in Partamies et al. (2017). It is based on visual viewing of the keogram (quicklook) data (available at http://www.space.fmi.fi/MIRACLE/ASC/). Faint stripy and patchy appearance of auroral emission in the ASC keogram is a signature of auroral pulsations. Due to the relatively low time resolution of the quicklook data, not all PsA events have been detected. We have rather selected periods where the pulsating structures cover most of the single station FoV. This was originally done to facilitate further analysis of auroral peak emission heights by using an automatic method introduced by Whiter et al. (2013). These height data are utilized in the current study as well. The height method is a modern version of triangulation assigning one height value for each image pair. An image pair means any two simultaneous images from neighboring ASC stations with overlapping FoVs. The brightness distributions from the two images are then mapped both to a horizontal and field-aligned plane. The peak emission height values are saved for images, for which the two mappings agree. The method was validated with synthetic data and tested with real images for which the peak emission heights had previously been triangulated manually. We use the height estimates for the green line (557.7 nm) image, because the green image cadence is higher, but the method itself resolved equally well the prompt emission at 427.8 nm. The method is capable of resolving changes in peak emission heights of less than or about 1 km. Variations in the peak emission height along individual auroral structures have been observed to be an order of magnitude larger (Sangalli et al., 2011). Because the PsA events were selected to fill most of the FoV, the corresponding peak emission height value describes the average height of the pulsating structures in an image. Many PsA periods were observed to start during substorm activity. Substorm aurora is typically much more intense than PsA, and in case of a mixture of intense and faint aurora, the height method most often assigns the height of the most intense emission to the whole image. We therefore require the PsA to be the dominant feature in the analyzed images.

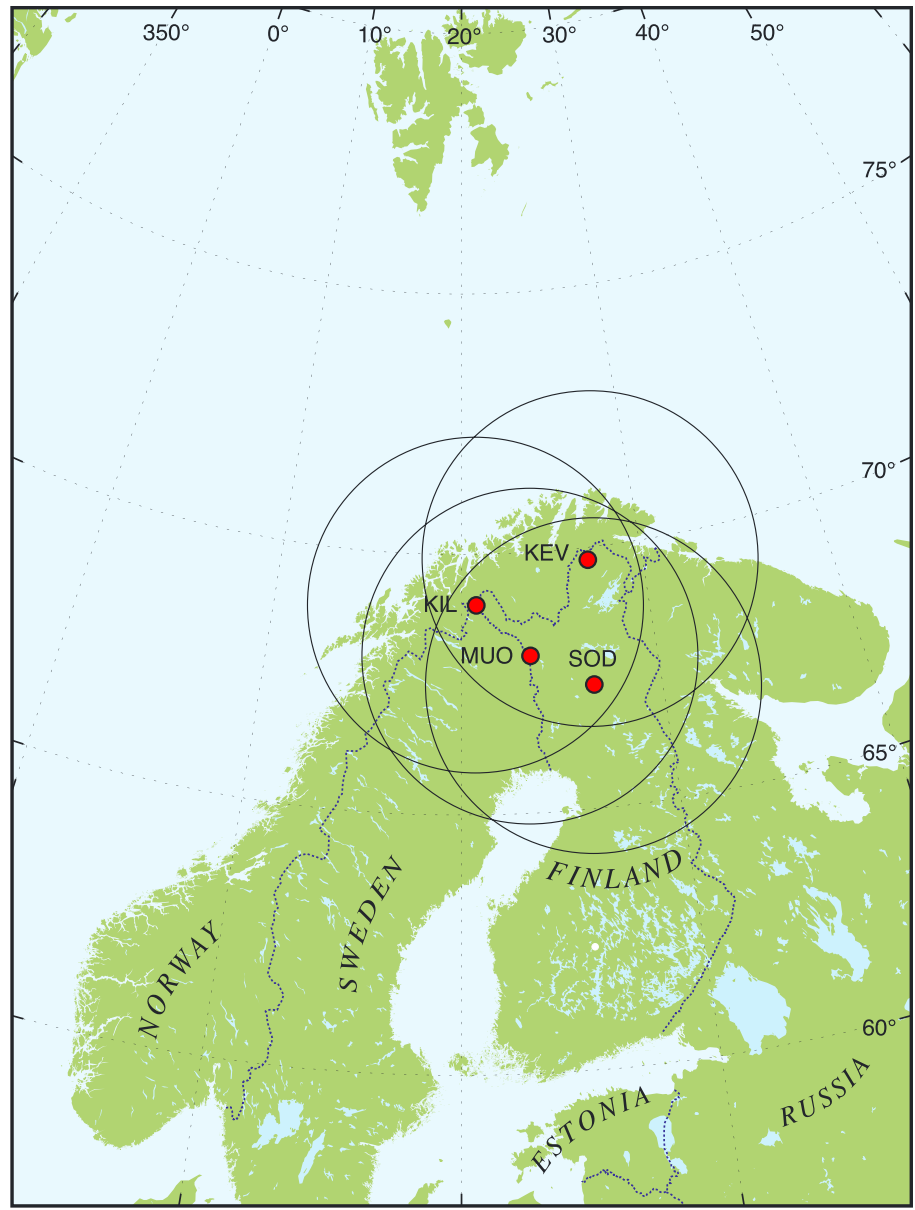

**Figure 1.** Station locations of the five MIRACLE all-sky cameras (red dots) used in this study. The fields of view are marked with circles around the stations.

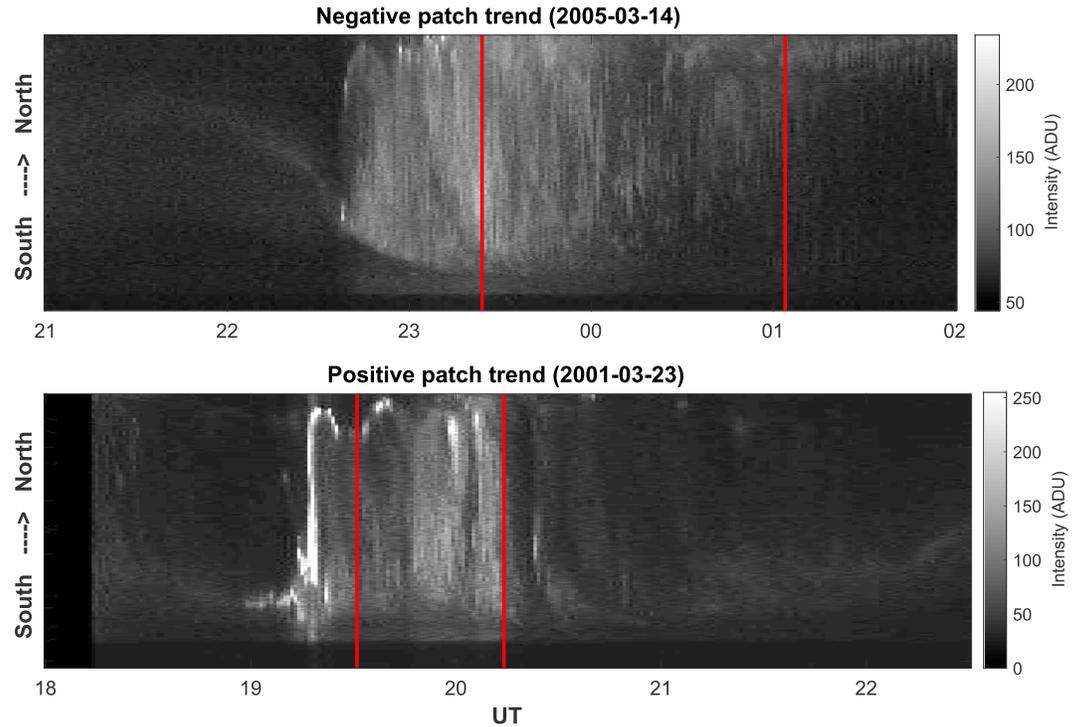

**Figure 2.** Keograms from MUO and SOD cameras showing examples of events with decreasing (top panel, 14 March 2005) and increasing patch size trends (bottom panel, 23 March 2001). Periods of pulsating aurora events are bracketed by vertical red lines. In case of decreasing patch size (top), the patchy features in the keogram become more sparse, while during the event with an increasing patch size trend (bottom), the patchiness of the keogram display becomes denser as a function of event time. Magnetic midnight in Lapland is at about 22 UT.

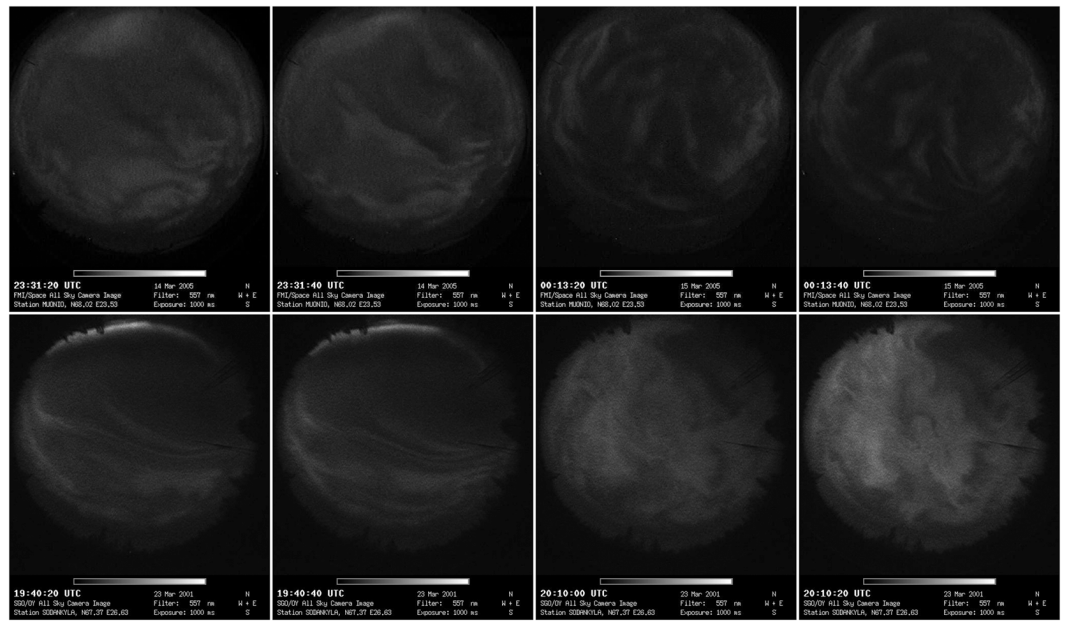

**Figure 3.** (top row) Consecutive MUO image pairs from the beginning (23:31:20 and 23:31:40 UT) and end of the event (00:13:20 and 00:13:40 UT) with a decreasing patch size trend. (bottom row) Consecutive SOD image pairs from the beginning (19:40:20 and 19:40:40 UT) and end of the event (20:10:00 and 20:10:20 UT) with an increasing patch size trend.

Examples of PsA events with increasing and decreasing evolution of image-averaged patch size are shown in keogram data in Figure 2. Keograms are from Muonio (MUO, Glat 68.02°, CGMlat 64.72°) and Sodankylä (SOD, Glat 67.37°, CGMlat 63.92°) stations, respectively. The keograms are constructed from images with 1-min cadence, which is a long time compared to the individual auroral pulsations. Our identification method, however, benefits from the fact that many different pulsation frequencies take place simultaneously resulting in a patchy display of diffuse aurora at any given moment. The periods of PsA events are bracketed by vertical red lines. The event with decreasing average patch size (top panel) took place during the substorm recovery phase and continued until the end of the recovery of the magnetic disturbance. The event ended when the observable patches moved poleward and became a small part of the FoV. In these cases, PsA can sometimes be seen north of Lapland, over Svalbard camera stations later in the morning (Partamies et al., 2017). In the case of increasing average patch size (bottom panel), the event took place during the substorm expansion and ended when PsA became overtaken by a weak nonpulsating diffuse emission band at the southern part of FoV during the recovery phase of the substorm. This event lasted for a shorter time than the event with decreasing patch sizes. Consecutive sample images from the beginning and end of the period of PsA for events in Figure 2 are shown in Figure 3. The top row consists of MUO images at 23:31:20 and 23:31:40 UT at the beginning of the event and those at 00:13:20 and 00:13:40 UT at the end of the event. In the first two images the regions turning dim and bright are large compared to those in the last two images. The bottom row includes SOD images at 19:40:20 and 19:40:40 UT at the beginning of the event and images at 20:10:00 and 20:10:20 UT at the end of the event. In the first two images the pulsations appear in thin and elongated structures, while in the last two images very large regions turn dim and bright.

The evolution of the local electrojet indices (IL and IU) for the events in Figure 2 are shown in Figure 4. The calculations of ionospheric (blue) and telluric (red) current contribution to the indices (black) based on the IMAGE magnetometer chain data (10-s resolution from 35 stations across the Fennoscandia and Svalbard) are described by Juusola et al. (2016). As this new version of the method for calculating ionospheric equivalent currents and the electrojet indices includes the contribution of the telluric currents, it provides the ionospheric contribution with an improved accuracy. The beginning and end of the PsA events are marked by the vertical lines. In the top panel, the event with decreasing average patch size is seen to start at around the maximum magnetic deflection and continue close to the end of the magnetic recovery. The event with increasing average patch size (bottom panel), however, started earlier during the substorm activity and lasted barely past the maximum magnetic deflection. The given examples illustrate the typical relative timings of the two subgroups of PsA with respect to the substorm activity.

## 3. Detecting Patch Sizes and Patch Size Trends

We use a recently developed method to detect patches and to follow the temporal evolution of patch sizes as seen in auroral images (Bolmgren, 2017). In order to locate pulsating patches, a nonpulsating background is modeled for each ASC image in the PsA event using two different approaches. The nonpulsating background changes slowly due to features such as the moonlight. The difference between the background and the original ASC image then reveals the patches at their "on" state in each time step. Prior to the background subtraction, the ASC images are preprocessed. The instrument noise (dark current) is removed by subtracting an average number of counts sampled from the black corners of each image. The pixels below the 60° zenith angle are cropped out to avoid large distortions due to fish-eye optics and the areas including unwanted objects close to the horizon, such as trees and buildings. The image data used in this study are collected at a 20-s cadence through an optical filter with a bandpass around 557.7 nm.

An illustration of the patch detection method by Bolmgren (2017) is included in Figure 5. An original raw ASC image from Figure 3 is used as an example (Figure 5a). The overlaid square shows the region that is cropped out for further analysis (zenith angle smaller than 60°). Three consecutive cropped images are shown in panels (b)–(d). A simple method to describe the nonpulsating background at a given time $t_i$ uses the previous and following images. The background is estimated using a point-by-point average of the previous and subsequent image at $t_{i-1}$ and $t_{i+1}$ (Figures 5b and 5d), which assumes that the images only experience a slow change. This averaged brightness is then median filtered using a 3 × 3 pixel window to smooth out small-scale variations. This approach and another more advanced spatiotemporal filtering method are both described and evaluated by Bolmgren (2017). It was concluded that the methods give very similar results for the data set of green auroral images at 20-s cadence. Thus, the simpler point-by-point average approach is used in this study.

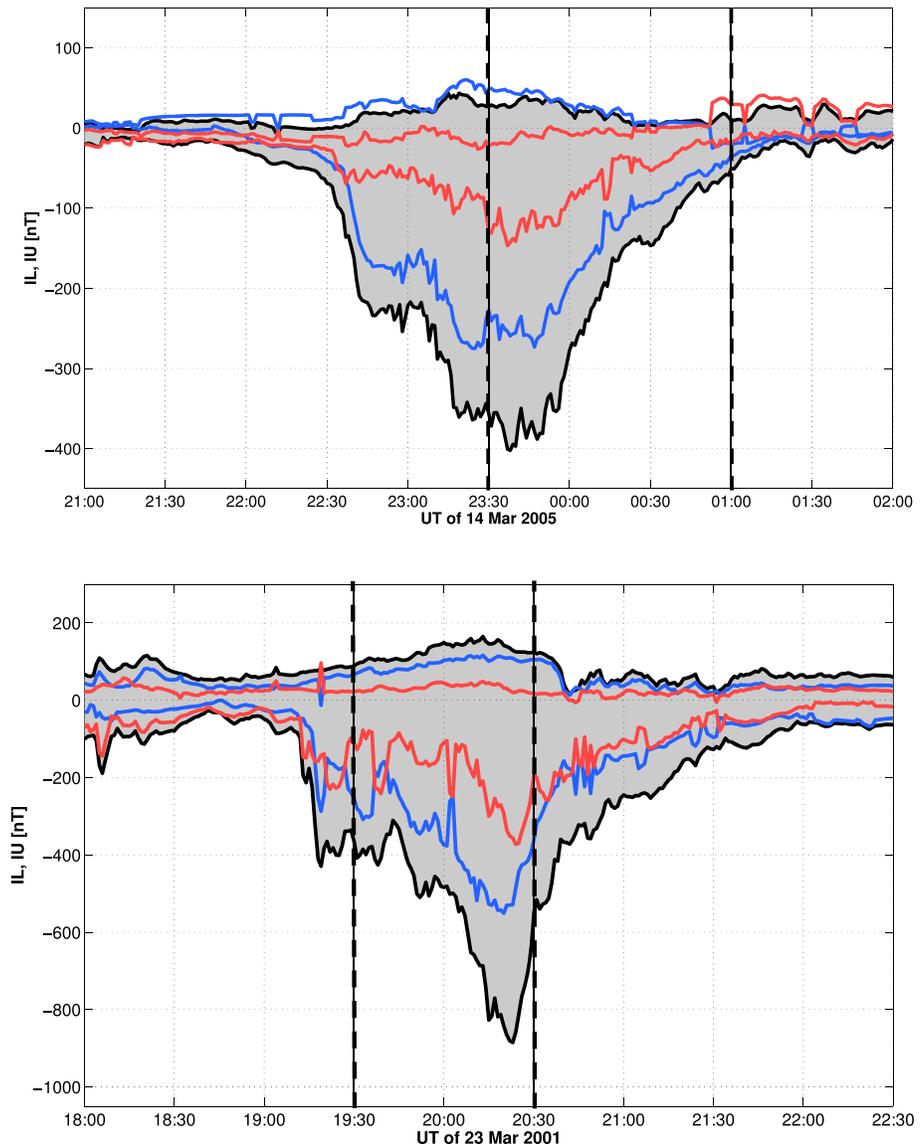

**Figure 4.** Local electrojet index evolution for sample events shown in Figure 2 with decreasing (top panel, 14 March 2005) and increasing patch size trends (bottom panel, 23 March 2001). Periods of pulsating aurora events are bracketed by vertical dashed lines. The blue curves describe the contribution of the ionospheric currents, and the red curves mark the contribution of telluric currents.

After subtracting the background (a smoothed average of Figures 5b and 5d) from the original images, we are left with the auroral structures (the foreground), which are patches during the PsA events (Figure 5e). Individual patches were required to contain 200 or more interconnected pixels (connected by sides or by corners) with brightness over an experimentally selected threshold value of 6 counts over the nonpulsating background.

Estimating the patch sizes in physical area requires an assumption for the height of auroral emission. Here, we used the height values calculated for these events in the previous study by Partamies et al. (2017) using the method of Whiter et al. (2013). The result is a size in square kilometers for each detected patch. Furthermore, two values are determined for each image: the total area of all foreground structures (patches) and the total foreground area divided by the number of patches, which gives an average patch size per image. The time series of these values indicate the temporal evolution of patch size during a PsA event without tracking any individual objects in the images. A linear fit is applied to each patch size time series. The sign and magnitude of the corresponding fit slope are used as indicators of patch size trend as a function of event time.

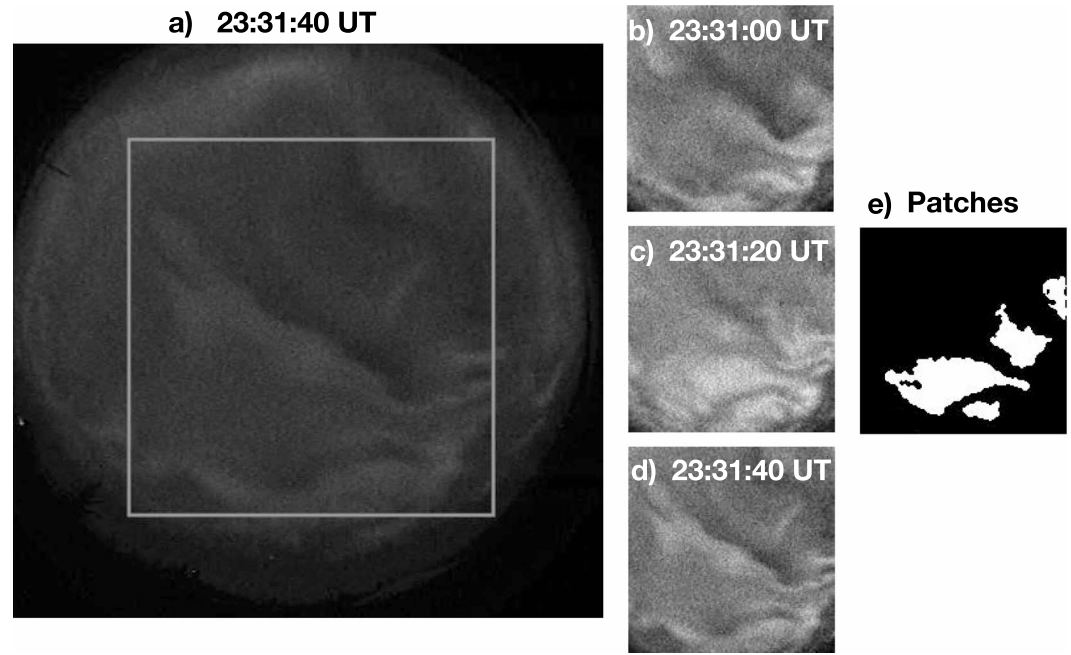

**Figure 5.** Sample images from the patch detection method: (a) a raw all-sky camera image at 23:31:40 UT (same as the second image in Figure 3) with the overlaid square showing the cropping area of the images during the preprocessing step, consecutive cropped images at 23:31:00 UT (b), 23:31:20 UT (c), and 23:31:40 UT (d) with dark current subtracted, and (d) patches (white, foreground) after subtracting the slowly evolving background (average of b and d) from the preprocessed image c). The patches detected at 23:31:20 UT (images c and e) are the ones which are "on" there but "off" during the exposures of the image before and after.

It should be noted that the image interval of 20 s is of the order of the longest periods typically observed in PsA (2–20 s). This means that only the individual patches with the longest periods are captured multiple times in a row, but the method rather follows the changes in the average patch size throughout the event lifetime. Bolmgren (2017) investigated the effects of this undersampling by comparing results of image data sets with cadences of 2 and 20 s. As the observed event-averaged trends in the patch size evolution were very similar, it was concluded that the patch size trends were not affected by the undersampling.

## 4. Events With Positive and Negative Patch Size Trends

The method was able to detect patch sizes for 148 events with good coverage throughout the event lifetime. This included 109 events with negative and 39 events with positive patch size trend. Since the deviations from the linear fit vary very much from event to event, suggesting that not all PsA experiences systematically decreasing or increasing patch size evolution, we have chosen a subset of events with the best linear fits (lowest standard deviation) as extreme examples of increasing and decreasing patch sizes. If these subsets of events do not show a different behavior, there is little point in analyzing events with more complex patch size evolution in more detail. The subsets include 35 events with decreasing and 18 events with increasing patch sizes, which we use for further analysis in the rest of the paper. Examples of linear regressions for decreasing (top) and increasing (bottom) patch sizes are shown in Figure 6. Both sample events experience a negative/positive trend in the total patch area (top panel) as well as the image-averaged patch size (middle panel). The number of patches per image (bottom panel) varies between 2 and 10 in both cases.

The combination of the average patch size and the number of patches causes the diffuse emission in the keograms to look denser or sparser. Because the parameters are difficult to untangle in the keogram data, our subsets are chosen so that the trends in both patch size parameters have the same sign. The total patch area describes the portion of the FOV, which is covered by patches. The distribution of this total patch area for events with decreasing (blue) and increasing (red) patch sizes is plotted in the left panel in Figure 7. The event-averaged total patch area varies mainly in the range of 400–3,000 km$^2$ for both subsets of events. During the events with a positive patch size trend the patches occupy slightly larger total area as compared to the events with a negative patch size trend. The average patch size (right panel in Figure 7) and the number of patches typically experience more variability than the total patch area does. This is due to the fact that the detection method selects individual patches based on a simple difference. The set of selected patches may vary from one time step to another, depending on the ongoing pulsation frequencies. The image-averaged patch size for all events in the data set generally stays within 100 to 800 $km^2$ with a typical value of about 350 $km^2$ for both subsets. The peak size for the events with decreasing patch sizes (blue) is slightly larger than that for the events with increasing patch size trends, but as the selection process is partly determined by the temporal resolution of the data, we will not draw any detailed conclusions based on these numbers.

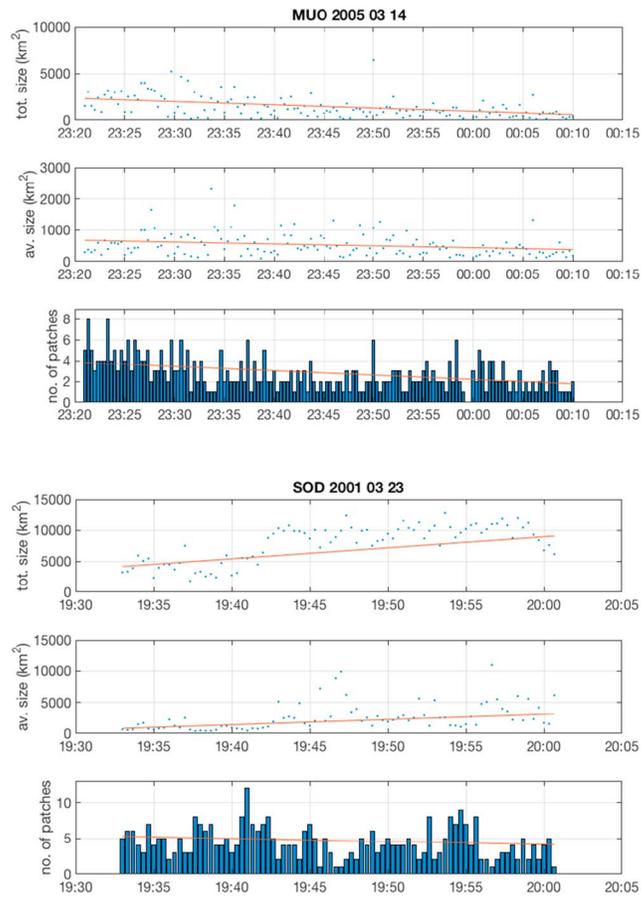

**Figure 6.** Linear regressions of the patch size evolution as a function of event time for the two events shown in Figure 2. A decreasing (top) and increasing (bottom) linear fit slopes follow the data points well, despite the variation from image to image. Standard deviation of the averaged patch size per image (central panels) is 360 km$^2$ in the decreasing event and 2,200 km$^2$ in the increasing event. For the total patch size per image (top panels) the corresponding standard deviations are 1,140 and 2,940 km$^2$. The standard deviation for the number of patches in each frame is 1.6 for the decreasing event and 2.6 for the increasing event.

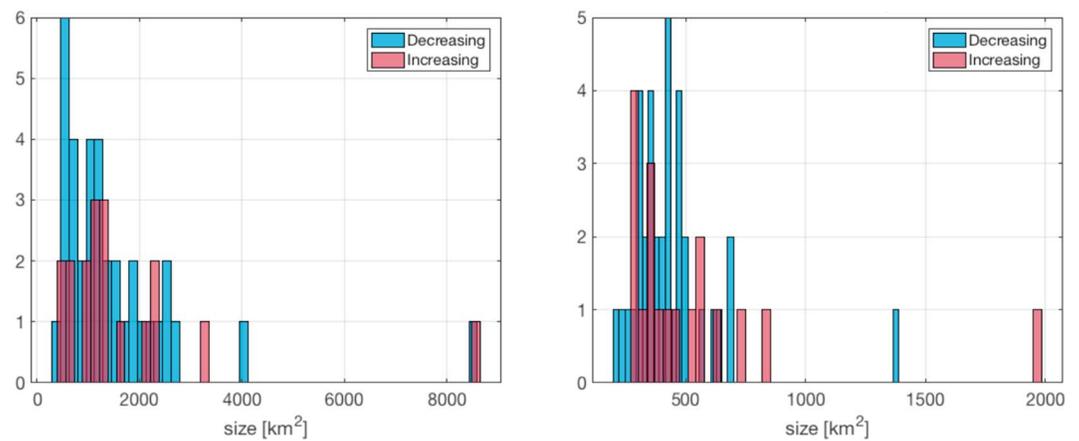

**Figure 7.** (left) The distribution of the event-averaged total patch area for the events with decreasing (blue) and increasing (red) patch sizes. (right) The distribution of the median image-averaged patch size per event for the events with decreasing (blue) and increasing (red) patch sizes.

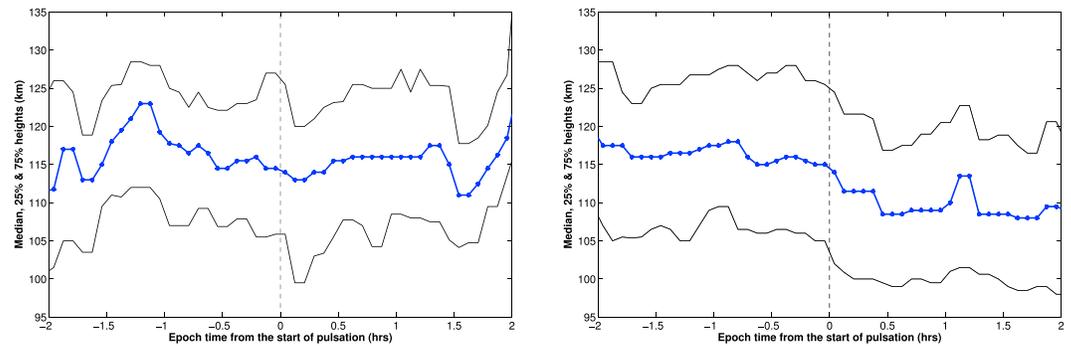

**Figure 8.** The superposed epoch analysis of the peak emission heights for events with positive (left) and negative (right) patch size trends. The zero epoch (vertical dashed line) marks the beginning of the pulsating aurora event. The epoch median values are shown by blue lines, and the 25th and 75th percentiles are plotted in black. The peak emission heights for the events with increasing patch size trends (left) do not experience a significant change when the pulsating aurora starts.

A number of parameters have been analyzed for the two subsets of PsA events to see if any significant differences can be observed. Figure 8 shows the superposed epoch analysis of the auroral peak emission height evolution for the events with increasing (left) and decreasing (right) patch sizes. The blue curves present the median height values, while the black curves are used for the evolution of the 25th and 75th percentiles. The zero epoch is the beginning of the PsA events. The displayed data set is a subset of the heights analyzed in Partamies et al. (2017). The median peak emission height for the events with positive trends (left) shows no significant changes during the PsA event, but the height varies around 115 km over the course of ±1 hr from the start of the event. In contrast, the median peak emission height for the events with negative trends (right) decreases by about 6 km during the first half an hour of PsA. This evolution is very similar to the median evolution of the full event set analyzed by Partamies et al. (2017), where they found a median peak emission height decrease of about 8 km during the first half an hour of PsA. Although the auroral emission originates in the *E* region, mainly above 100 km, the systematic peak emission height decrease can be used as a proxy for energetic electron precipitation which reaches the *D* region ionosphere.

Figure 9 displays CNA variation at about 30 MHz for the two subgroups of events. CNA data are from wide beam (60°) La Jolla riometers maintained and operated by the Sodankylä Geophysical Observatory (SGO) of the University of Oulu (Finland). The riometer station closest to the ASC stations have been used. For ASC stations at ABK and SOD, the two instruments are colocated, but for the ASC at MUO the Sodankylä riometer is the closest, for the ASC at KIL the closest riometer is at Abisko, and for the ASC at KEV the Ivalo riometer is used. The absorption evolution during individual events (gray) shows considerable event to event variation with CNA values between 0 and 5 dB, and consequently, the epoch median values are only marginally different between the two subgroups. The most pronounced difference is the absorption evolution during the epoch time 1–2 hr. The events with negatives patch size trends show a steady absorption around 0.5 dB until about 1.5-hr epoch time and thereafter a gentle decrease with decaying PsA (right panel). Most of our events have a lifetime shorter than 1.5 hr. The steady evolution of absorption is expected, since fluxes of energetic electrons during PsA are known to be low. The hard precipitation starts already during the substorm and remains at a stably elevated level throughout the following pulsation event, even when the softer precipitation decays. The events with positive patch size trends experience a mild increase of the median CNA (left panel). Although very subtle, the difference in the CNA signatures agrees with our earlier observation in that the PsA with positive patch size trends are taken over by substorm-type aurora at the end of the event lifetime. The substorm aurora means higher particle fluxes at all energies, which increases CNA from the PsA level.

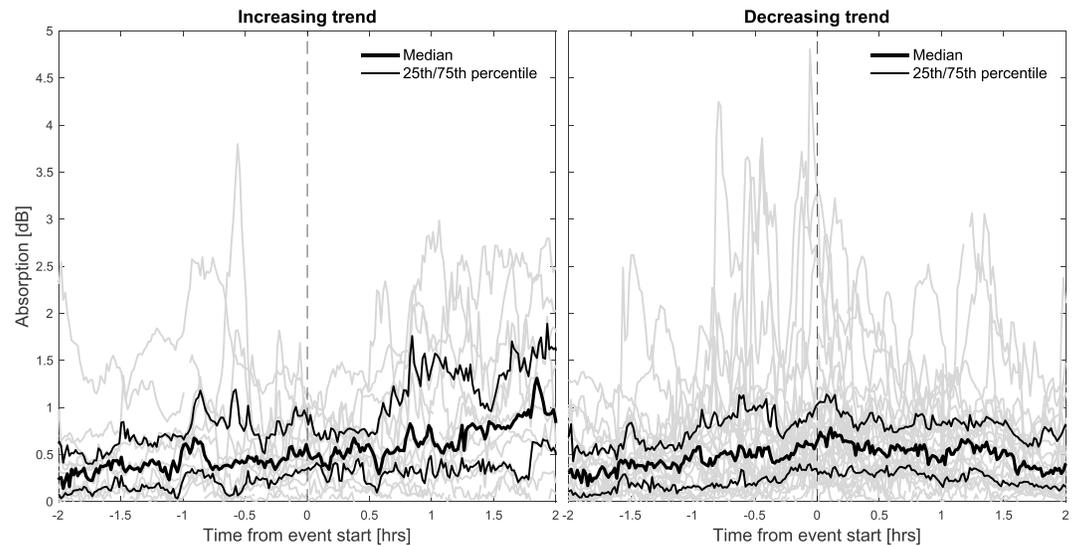

**Figure 9.** The superposed epoch of absorption for the events with a increasing (left) and decreasing (right) patch size evolution. The individual event curves (gray) and epoch median (black) are displayed. The dashed line marks the zero epoch for the beginning of the pulsating aurora events and the thin black curves show the 25th and 75th percentiles.

The patch size trends (slopes of the linear regression) detected in our set of events do not generally correlate with any magnetospheric, ionospheric, solar wind parameter, or the average patch sizes. However, the subgroup of events with decreasing patch sizes shows a mild correlation (Pearson correlation coefficient of 0.53) between the event duration and the total patch area trend. A relationship of a stronger positive correlation (0.70) for events with durations less than about 1.5 hr can be seen in Figure 10. This means that

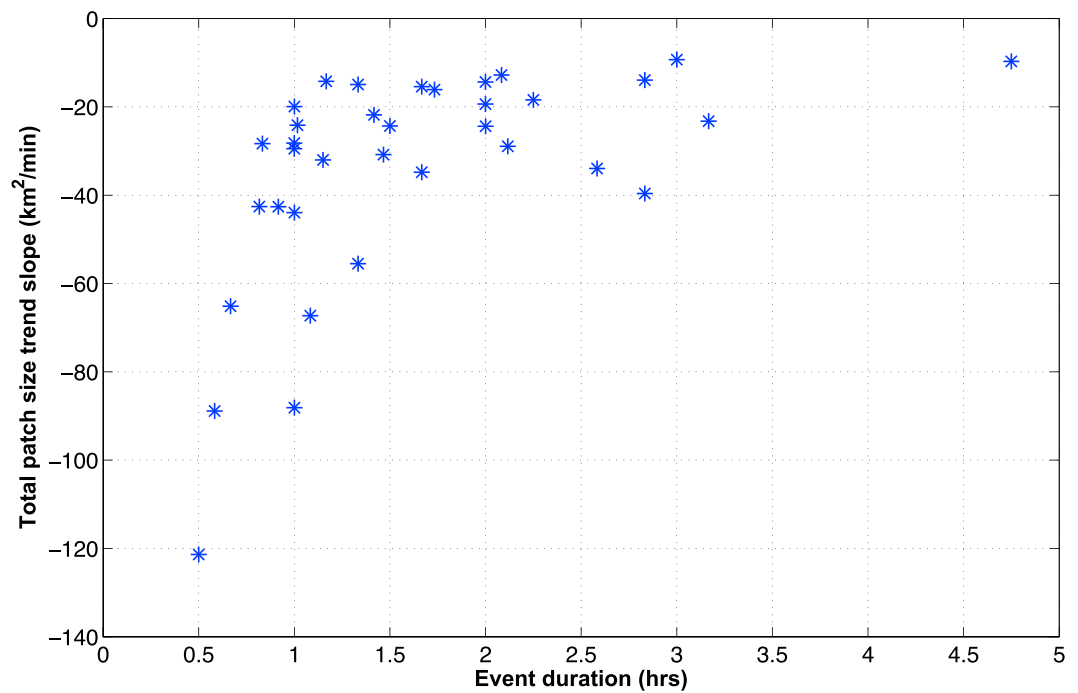

**Figure 10.** The relationship between the total patch area trend and the lifetime of the pulsating aurora events with decreasing patch size trends experiences a mild negative correlation.

**Table 1**
*A Selection of Geomagnetic Parameters for the Two Subsets of Pulsating Aurora Events*

| Parameter | Shrinking patches (35)<br>25% / median / 75% | Growing patches (18)<br>25% / median / 75% |
| --- | --- | --- |
| Event MLT | 02:12 / 03:50 / 04:18 | 02:36 / 03:20 / 03:48 |
| Dst index (nT) | −41 / −29 / −24 | −41 / −30 / −26 |
| AL index (nT) | −340 / −230 / −120 | −380 / −290 / −160 |
| $IL_{ASC}$ index (nT) | −310 / −170 / −70 | −340 / −240 / −110 |
| Peak emission height (km) | 99 / 109 / 119 | 104 / 114 / 124 |
| Solar wind speed (km/s) | 408 / 510 / 579 | 414 / 550 / 636 |
| IMF (nT) | 5.8 / 7.3 / 9.4 | 5.4 / 6.5 / 10 |
| IMF $B_Z$ (nT) | −4.2 / −1.6 / 0.9 | −4.1 / −2.4 / −0.4 |
| IMF $B_X$ (nT) | −4.1 / −1.0 / 3.8 | −3.5 / 1.2 / 3.4 |
| Absorption (dB) | 0.15 / 0.37 / 0.73 | 0.08 / 0.45 / 0.78 |
| $F_{130}$ | $1.6 \cdot 10^4$ / $2.7 \cdot 10^4$ / $5.0 \cdot 10^4$ | $1.6 \cdot 10^4$ / $5.2 \cdot 10^4$ / $7.5 \cdot 10^4$ |
| $F_{60}$ | $1.5 \cdot 10^5$ / $2.4 \cdot 10^5$ / $4.6 \cdot 10^5$ | $1.7 \cdot 10^5$ / $5.0 \cdot 10^5$ / $7.7 \cdot 10^5$ |

*Note.* The values are given as median values as well as 25th and 75th percentiles over the event lifetimes.

the shorter the event lifetime, the faster the patch evolution is from denser to sparser. For PsA events lasting longer than 2 hr a "saturation" level is reached and the trend in the total patch area fluctuates around a value of about −20 km². Corresponding, if much milder (−0.37) negative correlation may be present for the events with increasing patch size trends (data not shown), but unless much more events can be collected and successfully fitted, the significance of that result cannot be judged.

Table 1 shows a comparison of a set of indices and parameters for the two subsets of PsA events with decreasing and increasing patch sizes (shrinking and growing, respectively). The values are calculated as median, 25th and 75th percentile values over the event durations. The magnetic local time of the events does not show a significant difference between the two subgroups. The global geomagnetic index values (Dst and AL) are also comparable between the two subgroups, while the local electrojet index ($IL_{ASC}$ which is based on the magnetic measurements across the Lapland camera stations; Partamies et al., 2015) is 90 nT more negative for the events with growing patch sizes. The last two rows of the table include the indices that measure the global intensity of substorm-injected energetic electrons around the geosynchronous orbit (67° Mlat) at 60 and 130 keV (Borovsky & Yakymenko, 2017b). These indices are constructed from the electron flux measurements by multiple spacecraft as the largest 6-min median flux per hour. The chosen energies describe the radiation belt seed electron population, which is provided by the substorm injections and prone to wave-particle interaction. In agreement with the IL values the radiation belt electron indices are roughly twice as high for the cases of growing patches than those of shrinking patches. These findings further support the fact that the growing patches occur earlier during the substorm activity when the magnetic activity is still elevated and the inner magnetosphere still has a significant storage of energetic particles. The solar wind speed and the interplanetary magnetic field (IMF) magnitude are not significantly different in the two subgroups. IMF $B_Z$ is only slightly more negative for the events with growing patches, while the IMF X-component is typically negative for the events with decreasing and positive for the events with increasing patch sizes. The negative IMF X-component occurs in the away sector of the solar wind and the positive X-component in the toward sector of the solar wind. Thus, this suggests that the solar wind energy transfer may control the patch size evolution of the PsA and/or there is a strong seasonal bias between the two PsA subgroups. In fact, nearly 70% of the events with increasing patch sizes occur during the spring months (January–March) when the positive IMF $B_X$ provides a more efficient energy transfer from the solar wind to the magnetosphere (e.g., Borovsky & Yakymenko, 2017a). About 40% of the events with decreasing patch sizes take place during the autumn months (September–November), but in addition, there is about an equally strong occurrence maximum in December and around the spring equinox.

## 5. Discussion

Little is known about the factors controlling or relating to the sizes of the PsA patches. Earlier studies such as Samara and Michell (2010) show that the patch size correlates with the pulsation frequency. Larger patches experience lower-frequency pulsations as compared to smaller patches. The pulsation frequency, on the other hand, is driven by the wave-particle interaction in the magnetospheric source region. Thus, the spatial patch evolution may directly relate to the spatial coherence in the behavior of the chorus wave elements, as recently suggested by Ozaki et al. (2018). What exactly controls the coherent behavior of chorus elements, and the precipitation energy, requires more detailed future studies.

The patch size evolution was discussed by Partamies et al. (2017), as in whether the changes between arc-like and patch-like pulsating structures or the variation in the patch size in general could be related to changes in the particle precipitation energy. A sample event in that study illustrated a decay in the structure size and a simultaneous decay in the total electron energy flux in the auroral energy range (<8 keV). The decay of the pulsation structure size was speculated to be a common feature after substorm activity and based on our findings here that seems to be the most likely scenario. The fact that the PsA continues and the mild CNA signature persists while the energy flux (and magnetic and precipitation indices) decay suggests that the decay is primarily due to decreasing fluxes of low-energy precipitation while the fluxes of higher energy particles still remain elevated.

Selected patches (rather than all patches) have been detected by the method used in this study. However, it is likely that the selected patches represent the patch population with the longest lifetimes, partly owing to the low temporal resolution of the data. It is, of course, possible that more transient patches with different size evolution coexist but have not been detected. The most stable patches are typically the ones leaving the trace in the keograms, which have 1-min temporal resolution. As this method is compatible with the keogram data, it helps in separating the events into different energy ranges, and it can be a useful tool in the future studies. In our selected events, the patch size either increase or decrease systematically enough to allow linearly fitted trend line. In the full trend data set more complicated evolutions were observed as well. Whether those mean variable energy input into the atmosphere is beyond what can be concluded without directly measuring the incoming particles. For more detailed studies a higher temporal resolution of the camera data would also be beneficial.

A majority of the PsA (about 60%) has been related to substorm recovery phases before (Lessard, 2012; Partamies et al., 2017). The current study suggests that the events occurring during the substorm expansion phases may be tightly embedded into the substorm aurora and may thus not add much to the substorm-related energy deposition of energetic electron precipitation. A bigger question is the effect of the PsA events with decreasing patch size trends. During those events the magnetic disturbances of the substorms have already recovered while the energetic precipitation can have a much longer lifetime. The decay of high fluxes of soft precipitation toward the end of the substorm recovery phase and the late timing of PsA during substorms together explain why the peak emission height evolution does not correlate with magnetic indices (Partamies et al., 2017). While the decrease in the peak emission height indicates increased precipitation energies (from a few kiloelectron volts to about 10 keV), the magnetic indices are more sensitive to high fluxes of softer particles (<8 keV), which mainly contribute to the auroral brightness and ionospheric currents. Despite the low flux values, the high-energy precipitation component may have an important effect on the mesospheric chemistry (Turunen et al., 2016).

Structural changes in the PsA is an unresolved topic, largely because it is technically challenging to detect individual pulsating patches and to determine and track their sizes or shapes throughout their evolution. Tracking methods used by Grono et al. (2017) and Humberset et al. (2016) produce good estimates of the patch propagation, but the patch detection and tracking is limited to stable structures. More recently, Humberset et al. (2018) investigated the morphology of four individual patches. They reported patch sizes of 1,000–5,000 km$^2$ assuming an emission height of 110 km (horizontal scale sizes of some 10 km). These numbers of individual events roughly agree with our results of patch sizes typically between about 200 and 2,000 km$^2$ at heights of 110–115 km. The individually tracked patches are a sample of a size spectrum for more stable or long-lived patches. The lifetimes of individual patches analyzed by Humberset et al. (2018) were indeed 7–10 min. Furthermore, three of their four analyzed patches keep at least 85% of their shape during their lifetime. These kind of patches will leave a trace in the 1-min keogram data. They are also likely to be captured by our patch detection as well, although our method only requires the patches to be in an "on" state during the exposure.

Another recent study by Grono and Donovan (2018) investigated the morphology of PsA in more detail. They reported the first categorization of PsA structures into patchy aurora (PA), patchy pulsating aurora (PPA), and amorphous pulsating aurora (APA). According to this classification, pulsations in PA are limited to small regions, while those of PPA involve most of the patch area. APA is classified as very unstable, rapidly changing and basically untraceable. Our events would mainly fall into the stable pulsation category of PPA, while PA would mostly likely be seen as slowly changing (drifting) background. At the beginning of the events with negative patch size trend or at the end of the events with positive patch size trend some APA may be included, if sufficiently large emission regions would be "on" at the time of the exposure. The evolution of the patch size or number of patches was not examined in the Grono and Donovan (2018) paper but would make an interesting future study.

Patch size detection in this study provides a longer-term overview of patchiness at any given time during the event without tracking individual structures. Our results suggest that the energy deposition due to PsA, which is not described by magnetic activity, is associated with the PsA events with shrinking patches. We have focused on the analysis of events with clear increase or decrease in the patch size, as the two extreme cases. More complicated patch size evolution was also observed, but detailed analysis of that would require image data with higher temporal resolution, as well as a proper electron precipitation measurement rather than a proxy. Fast imaging together with, for instance, radar measurements of electron density can shed the light on these details in the future.

## 6. Conclusions

We have used a simple approach to detect pulsating patches in ASC images for 148 periods of PsA. Detected patches were analyzed further to estimate the patch sizes for every image, both in image-averaged patch size and in total patch area per image. We further selected 53 events where the average patch size and the total patch area coherently increase (18) or decrease (35) during the event lifetime. An analysis of the geomagnetic conditions for the two subsets of PsA showed small but consistent behavior. The events with increasing patch sizes took take place mainly during substorm expansions (elevated IL). During these events the peak emission height of the aurora is not changing. The indices describing the substorm-injected energetic electrons had twice as high values during the event with increasing patch sizes as compared to those with decreasing patch sizes. The events with decreasing patch sizes occur in the substorm recovery phase (lower IL and radiation belt flux index values). During these events the peak emission height decreases by about 6 km. The patch size itself is a very variable parameter and does not show significant event-averaged differences.

We further found that the events with decreasing patch sizes typically last for about 15 min longer than those with increasing patch sizes. For the events with decreasing patch sizes, an anticorrelation was observed between the patch size change rate and the event duration: the faster the patch size is decreasing, the shorter the event is, on average. For PsA events longer than 1.5 hr, the patch size decay rate varies around $-10$ to $-40$ km$^2$/min.

While it would be important to study the patch size evolution in more detail and also include the events with alternating patch size trends, our results suggest that the part of the energy deposition, which is not captured by the substorm evolution and magnetic indices, is the one with decreasing patch size trend. This can be an important visual indicator for events, which may have a significant impact on the mesospheric chemistry.

**Acknowledgments**

This work was supported by the Research Council of Norway under CoE contract 223252. Index data from Kyoto World Data Center, and solar wind data from OMNIWeb. The authors thank K Kauristie, S. Mäkinen, J. Mattanen, A. Ketola, and C.-F. Enell for careful maintenance of the camera network and data flow. MIRACLE ASC quicklook data are available at http://www.space.fmi.fi/MIRACLE/ASC/asc_keograms_00.shtml and at GAIA virtual observatory at http://gaia-vxo.org, while full-resolution image data can be requested from FMI (kirsti.kauristie@fmi.fi). Sodankylä Geophysical Observatory riometer data quicklook plots can be browsed at http://www.sgo.fi/Data/Riometer/rioData.php and the data are available upon request. Radiation belt flux index data are available upon request from J. Borovsky.